\documentclass[final]{elsart}
\usepackage{graphicx}
\usepackage{amsmath}
\usepackage{amssymb}
\usepackage{mycommands}
\usepackage{subfigure}
\usepackage{epsfig}
\usepackage{lscape}

\journal{Physics Letters B}

\def\nm {\mbox{\boldmath $\nu_\mu$}} 
 
\def\ne {\mbox{\boldmath $\nu_e$}}

\def\nmne{\mbox{\boldmath $\nu_{\mu}\rightarrow\nu_{e}$}}

\def\cohp{\boldmath {Coh$\pi^0$} }
\def\piz{\boldmath {$\pi^0 $} }

\def\gam{\boldmath {$\gamma $} }

\def\A{\boldmath {${\cal A}$} }

\def\ztg{\boldmath {$\zeta_{\gamma }$}}

\def\gam{\boldmath   {  ${ {  \gamma }} $} }
\def\sgam{\boldmath {1${ {  \gamma }} $} }
\def\tgam{\boldmath  {2${ {  \gamma }} $} }

\def\sv{\boldmath {${\rm V}^0 $} }
\def\dv{\boldmath {2${\rm V}^0 $} }
\def\pan{\boldmath {${\rm {PAN }}$ }}

\begin{document}

\begin{frontmatter}

\title{
A Search for  Single Photon Events in Neutrino  Interactions 
in NOMAD
}

\author[19]{C.T.~Kullenberg}
\author[19]{S.R.~Mishra}
\author[19]{D.~Dimmery}
\author[19]{X.C.~Tian}
\author[8]{D.~Autiero}
\author[8,12]{S.~Gninenko}
\author[8,24]{A.~Rubbia}
\author[25]{S.~Alekhin}
\author[14]{P.~Astier}
\author[18]{A.~Baldisseri}
\author[13]{M.~Baldo-Ceolin}
\author[14]{M.~Banner}
\author[1]{G.~Bassompierre}
\author[9]{K.~Benslama}
\author[18]{N.~Besson}
\author[8,9]{I.~Bird}
\author[2]{B.~Blumenfeld}
\author[13]{F.~Bobisut}
\author[18]{J.~Bouchez}
\author[20]{S.~Boyd\thanksref{Now1}}
\thanks[Now1]{Now at University of Warwick, UK}
\author[3,24]{A.~Bueno}
\author[6]{S.~Bunyatov}
\author[8]{L.~Camilleri}
\author[10]{A.~Cardini}
\author[15]{P.W.~Cattaneo}
\author[16]{V.~Cavasinni}
\author[8,22]{A.~Cervera-Villanueva}
\author[11]{R.~Challis}
\author[6]{A.~Chukanov}
\author[13]{G.~Collazuol}
\author[8,21]{G.~Conforto \thanksref{Deceased}}
\thanks[Deceased]{Deceased}
\author[15]{C.~Conta}
\author[13]{M.~Contalbrigo}
\author[10]{R.~Cousins}
\author[9]{H.~Degaudenzi}
\author[8,16]{A.~De~Santo}
\author[16]{T.~Del~Prete}
\author[8]{L.~Di~Lella \thanksref{Now2}}
\thanks[Now2]{Now at Scuola Normale Superiore, Pisa, Italy}
\author[8]{E.~do~Couto~e~Silva}
\author[14]{J.~Dumarchez}
\author[20]{M.~Ellis\thanksref{Now3}}
\thanks[Now3]{Now at Brunel University, Australia}
\author[3]{G.J.~Feldman}
\author[15]{R.~Ferrari}
\author[8]{D.~Ferr\`ere}
\author[16]{V.~Flaminio}
\author[15]{M.~Fraternali}
\author[1]{J.-M.~Gaillard}
\author[8,14]{E.~Gangler}
\author[5,8]{A.~Geiser}
\author[5]{D.~Geppert}
\author[13]{D.~Gibin}
\author[19]{A.~Godley}
\author[8,22]{J.-J.~Gomez-Cadenas}
\author[18]{J.~Gosset}
\author[5]{C.~G\"o\ss ling}
\author[1]{M.~Gouan\`ere}
\author[8]{A.~Grant}
\author[7]{G.~Graziani}
\author[13]{A.~Guglielmi}
\author[18]{C.~Hagner}
\author[22]{J.~Hernando}
\author[3]{P.~Hurst}
\author[11]{N.~Hyett}
\author[7]{E.~Iacopini}
\author[9]{C.~Joseph}
\author[9]{F.~Juget}
\author[11]{N.~Kent}
\author[6]{O.~Klimov}
\author[8]{J.~Kokkonen}
\author[12,15]{A.~Kovzelev}
\author[1,6]{A. Krasnoperov}
\author[19]{J.J.~Kim}
\author[12]{M.~Kirsanov}
\author[12]{S.~Kulagin}
\author[13]{S.~Lacaprara}
\author[14]{C.~Lachaud}
\author[23]{B.~Laki\'{c}}
\author[15]{A.~Lanza}
\author[4]{L.~La Rotonda}
\author[13]{M.~Laveder}
\author[14]{A.~Letessier-Selvon}
\author[14]{J.-M.~Levy}
\author[19]{J.~Ling} 
\author[8]{L.~Linssen}
\author[23]{A.~Ljubi\v{c}i\'{c}}
\author[2]{J.~Long}
\author[7]{A.~Lupi}
\author[6]{V.~Lyubushkin}
\author[7]{A.~Marchionni}
\author[21]{F.~Martelli}\
\author[18]{X.~M\'echain}
\author[1]{J.-P.~Mendiburu}
\author[18]{J.-P.~Meyer}
\author[13]{M.~Mezzetto}
\author[11]{G.F.~Moorhead}
\author[6]{D.~Naumov}
\author[1]{P.~N\'ed\'elec}
\author[6]{Yu.~Nefedov}
\author[9]{C.~Nguyen-Mau}
\author[17]{D.~Orestano}
\author[17]{F.~Pastore}
\author[20]{L.S.~Peak}
\author[21]{E.~Pennacchio}
\author[1]{H.~Pessard}
\author[19]{R.~Petti}
\author[8]{A.~Placci}
\author[15]{G.~Polesello}
\author[5]{D.~Pollmann}
\author[12]{A.~Polyarush}
\author[11]{C.~Poulsen}
\author[6,14]{B.~Popov}
\author[13]{L.~Rebuffi}
\author[24]{J.~Rico}
\author[5]{P.~Riemann}
\author[8,16]{C.~Roda}
\author[15]{F.~Salvatore}
\author[6]{O.~Samoylov}
\author[14]{K.~Schahmaneche}
\author[5,8]{B.~Schmidt}
\author[5]{T.~Schmidt}
\author[13]{A.~Sconza}
\author[19]{A.M.~Scott}
\author[19]{M.B.~Seaton}
\author[11]{M.~Sevior}
\author[1]{D.~Sillou}
\author[8,20]{F.J.P.~Soler}
\author[9]{G.~Sozzi}
\author[2,9]{D.~Steele}
\author[8]{U.~Stiegler}
\author[23]{M.~Stip\v{c}evi\'{c}}
\author[18]{Th.~Stolarczyk}
\author[9]{M.~Tareb-Reyes}
\author[11]{G.N.~Taylor}
\author[6]{V.~Tereshchenko}
\author[12]{A.~Toropin}
\author[14]{A.-M.~Touchard}
\author[8,11]{S.N.~Tovey}
\author[9]{M.-T.~Tran}
\author[8]{E.~Tsesmelis}
\author[20]{J.~Ulrichs}
\author[9]{L.~Vacavant}
\author[4]{M.~Valdata-Nappi\thanksref{Now4}}
\thanks[Now4]{Now at Univ. of Perugia and INFN, Perugia, Italy}
\author[6,10]{V.~Valuev}
\author[14]{F.~Vannucci}
\author[20]{K.E.~Varvell}
\author[21]{M.~Veltri}
\author[15]{V.~Vercesi}
\author[8]{G.~Vidal-Sitjes}
\author[9]{J.-M.~Vieira}
\author[10]{T.~Vinogradova}
\author[3,8]{F.V.~Weber}
\author[5]{T.~Weisse}
\author[8]{F.F.~Wilson}
\author[11]{L.J.~Winton}
\author[19]{Q.~Wu\thanksref{Now5}}
\thanks[Now5]{Now at Illinois Institute of Technology, USA}
\author[20]{B.D.~Yabsley}
\author[18]{H.~Zaccone}
\author[5]{K.~Zuber}
\author[13]{P.~Zuccon}

\address[1]{LAPP, Annecy, France}
\address[2]{Johns Hopkins Univ., Baltimore, MD, USA}
\address[3]{Harvard Univ., Cambridge, MA, USA}
\address[4]{Univ. of Calabria and INFN, Cosenza, Italy}
\address[5]{Dortmund Univ., Dortmund, Germany}
\address[6]{JINR, Dubna, Russia}
\address[7]{Univ. of Florence and INFN,  Florence, Italy}
\address[8]{CERN, Geneva, Switzerland}
\address[9]{University of Lausanne, Lausanne, Switzerland}
\address[10]{UCLA, Los Angeles, CA, USA}
\address[11]{University of Melbourne, Melbourne, Australia}
\address[12]{Inst. for Nuclear Research, INR Moscow, Russia}
\address[13]{Univ. of Padova and INFN, Padova, Italy}
\address[14]{LPNHE, Univ. of Paris VI and VII, Paris, France}
\address[15]{Univ. of Pavia and INFN, Pavia, Italy}
\address[16]{Univ. of Pisa and INFN, Pisa, Italy}
\address[17]{Roma Tre University and INFN, Rome, Italy}
\address[18]{DAPNIA, CEA Saclay, France}
\address[19]{Univ. of South Carolina, Columbia, SC, USA}
\address[20]{Univ. of Sydney, Sydney, Australia}
\address[21]{Univ. of Urbino, Urbino, and INFN Florence, Italy}
\address[22]{IFIC, Valencia, Spain}
\address[23]{Rudjer Bo\v{s}kovi\'{c} Institute, Zagreb, Croatia}
\address[24]{ETH Z\"urich, Z\"urich, Switzerland}
\address[25]{Inst. for High Energy Physics, 142281, Protvino, Moscow, Russia}

\begin{abstract}
We present a search for neutrino-induced events 
containing a single, exclusive photon  
using data from the NOMAD experiment at the CERN SPS 
where the average energy of the neutrino flux is $\simeq 25$ GeV. 
The search is motivated by an excess of electron-like 
events in the 200--475 MeV energy region as reported 
by the MiniBOONE experiment. 
In NOMAD, photons are identified via their 
conversion to $e^+e^-$ in an active target embedded in a 
magnetic field.  The background to the single photon 
signal is dominated by the asymmetric decay of  neutral pions  
produced either in a coherent neutrino-nucleus interaction,
or in a neutrino-nucleon neutral current deep inelastic scattering, 
or in an interaction occurring outside the fiducial volume. 
All three backgrounds are determined {\it in situ}  using control data samples  
prior to opening the `signal-box'. 
In the signal region, we observe {\bf 155} events with a predicted 
background of  {\bf 129.2 $\pm$ 8.5 $\pm$ 3.3}.  
We interpret this as null evidence for  excess of single photon 
events, and set a limit.   
Assuming that the hypothetical single photon has a 
momentum distribution similar to that of a photon from the 
coherent $\pi^0$ decay, the measurement 
yields an upper limit on  single photon events,  
{\boldmath $< 4.0 \times 10^{-4}$}  per  \nm\ charged current event. 
Narrowing the search to events 
where the photon is approximately collinear with the incident neutrino, 
we observe {\bf 78} events with a predicted 
background of  {\bf 76.6 $\pm$ 4.9 $\pm$ 1.9} yielding a more stringent  
upper limit,  {\boldmath $< 1.6 \times 10^{-4}$} 
per  \nm\ charged current event. 

\end{abstract}


\end{frontmatter}

\section{The Question}
\label{sec-intro}
The MiniBooNE experiment, in the neutrino-mode, 
has reported an  excess of $129 \pm 43$ (Stat $\oplus$Syst) 
electron-like events 
in the 200--475 MeV energy range~\cite{MB-EX}, a range largely unaffected by the 
\nmne\ oscillations at $\Delta m^2 > 1$ eV$^2$, which is 
expected to dominate the 475-1250 MeV range~\cite{MB-NMNE}. 
Assuming that the efficiency of these events is 0.25, 
similar to that of the \ne-induced electrons, the rate of the 
excess with respect to \nm-CC is $\simeq 3 \times 10^{-3}$. 
In the antineutrino mode, initially the MiniBOONE data 
were consistent with background in the  200--475 MeV range~\cite{MB-NMbNEb}.  
Recently, with additional  antineutrino data,  they 
report a  2$\sigma$ excess in this region~\cite{MB-NUFACT11}. 
MiniBOONE lacks the resolution to determine if the excess is  
due to   $e^-$, $e^+$, or  a photon. 
In neutrino interactions single photons do not occur:   
they occur in pairs  from the 
$\pi^0 \rightarrow \gamma \gamma$ decay. 
Evidence of  events with a single photon and nothing else 
in neutrino induced neutral current (NC) interaction will signal  either new 
or unconventional physics. 
This anomaly needs to be checked experimentally.  
The low-energy MiniBOONE excess has motivated novel 
hypotheses invoking conventional, to unconventional, to 
entirely new physics processes.  
An explanation~\cite{BODEK-BREM} 
of the excess attributed to the 
internal bremsstrahlung associated with  muons in 
\nm-induced charged current interaction (CC) has been 
refuted by MiniBooNE~\cite{MB-BREM}. 
Harvey, Hill, and Hill~\cite{3H-PRL, 3H-PRD} 
postulate a new anomaly-mediated interaction between 
the $\nu$-induced $Z^0$-boson, the photon, and a vector meson, 
such as $\omega$, coupling  strongly to the target nucleus. 
A similar, single photon process has been envisioned 
by Jenkins \& Goldman~\cite{NCRAD-JG}. 
In the anomaly-mediated neutrino-photon (ANP) interaction, 
$\nu {\cal N} \rightarrow \nu {\cal N} \gamma$, 
where ${\cal N}$ is the target nucleus or nucleon, 
the observable is a single \gam,  approximately 
collinear with the incident neutrino direction in the lab frame
with a recoiling ${\cal N}$ which remains intact and  largely undetected; 
the  form factor of ${\cal N}$ is expected to induce the 
single $\gamma$ to be pulled forward. 
The ANP cross-section is expected to increase with neutrino 
energy ($E_\nu$). The ANP interaction is akin to the exclusive $\pi^0$ production 
when the neutrino coherently scatters off a target nucleus (\cohp), 
{\boldmath $\nu + {\cal N} \rightarrow \nu + {\cal N} + \pi^0$}, 
where the only observable 
is $\pi^0 \rightarrow \gamma \gamma$.  
The phenomenology of ANP developed by Hill in~\cite{HILL09},  
and largely focussed on the low energy neutrino interaction
($E_\nu \simeq {\cal O}(1)$~GeV), predicts  a  
cross-section sufficiently large to explain the MiniBOONE anomaly. 
If true, the ANP interaction will also impact astrophysical processes 
such as neutron star cooling. 
Hypotheses involving new physics include: 
decay of a heavy neutrino into a light neutrino and a photon, 
$\nu_h \rightarrow \nu + \gamma$, by Gninenko~\cite{SG-NDKY, SG-NDKY10}; 
CP-violation in 3$\oplus$ 2 model by 
Maltoni \& Schwetz~\cite{CP32-MS}, and 
by Goldman {\it et al.}~\cite{CP32-GSM}; 
extra-dimensions in 3$\oplus$1 model by Pas {\it et al.}~\cite{ExDim-PPW}; 
and models by Giunti and Laveder~\cite{3p1-GL}; 
Lorentz-violation by Katori {\it et al.}~\cite{LV-KKT};
CPT-violation in 3$\oplus$1 model by Barger {\it et al.}~\cite{CPT31-BMW}; 
new gauge bosons with sterile neutrinos by Nelson \& Walsh~\cite{NGBSN-NW}; 
and soft-decoherence by Farzan {\it et al.}~\cite{SD-FSS}; etc. 
The NOMAD collaboration has reported a search of heavy-$\nu$ 
which mixes with  $\nu_\tau$'s  produced in the SPS proton-target, 
and then decays into $e^-e^+$ pair in the detector~\cite{NOMAD-HNU}. 
The search  focussed on highly collinear  
(${\cal {C}} = \left [ 1 - cos \Theta_{ee} \right ] \leq 2 \times10^{-5}$) 
and high energy ($E_{ee} \geq 4$~GeV) $e^-e^+$ events,  
where $\Theta_{ee}$ and $E_{ee}$ 
are the polar angle and the energy of the converted photon, 
and resulted in a single event consistent with the estimated background. 
The present work extends this search  with  no  cut on ${\cal C}$.

The high resolution NOMAD data allow a sensitive 
search for neutrino induced single photon events (\sgam). 
The detector measures  the energy and  emission angle of the photon. 
Additionally, the detector affords the redundancy to measure {\it in situ}  all  relevant 
backgrounds relieving the reliance on Monte Carlo simulation  (MC) 
of the conventional processes. As regards the MiniBooNE low-energy 
anomaly, we note that the average energy of the 
SPS neutrino flux at NOMAD, $E_{\nu} \simeq 25$~GeV, is 
much higher than that of   MiniBOONE, 
$E_{\nu} \simeq 1$~GeV.  
However, the NOMAD data provide a stringent check in a different 
energy range. If a mechanism explaining  
the anomaly were applicable to higher energies, such 
as  ANP or $\nu$-decay hypotheses, then 
this search, even if negative,  will furnish meaningful 
limits. 


\section{Beam and Detector}
\label{sec-nomad}

The Neutrino Oscillation MAgnetic 
Detector (NOMAD) experiment at CERN used a 
neutrino beam  produced by 
the 450~GeV protons from the 
Super Proton Synchrotron (SPS) incident on
 a beryllium target and producing  
secondary $\pi^{\pm}$, $K^{\pm}$, and $K^0$ mesons. 
The positively charged mesons were focussed by 
two magnetic horns into a 290~m long evacuated decay pipe. Decays of  
$\pi^{\pm}$,  $K^{\pm}$, and $K^0_L$  
produced the SPS neutrino beam. 
The average neutrino flight path to  NOMAD was 628~m, the detector being 
836~m downstream of the Be-target.  
The SPS beamline  and the neutrino flux incident 
at NOMAD are described in~\cite{NOMAD-FLUX}.  
The $\nu$-flux in NOMAD is constrained by the 
$\pi^{\pm}$ and $K^{\pm}$ production measurements in 
proton-Be collision by the SPY experiment 
~\cite{SPY1, SPY2, SPY3} and by an  earlier measurement conducted by 
Atherton {\it et al.}~\cite{ATHERTON}. The  $E_\nu$-integrated relative composition of 
\nm:\nmb:\nel:\neb\ CC events, constrained {\it in situ} by the 
measurement of CC-interactions of each of the  neutrino species,  is 
$1.00: 0.025: 0.015:0.0015$. Thus,  97.5\% of the events    
are induced by neutrinos with a small  anti-neutrino contamination, 
similar to that reported by MiniBOONE~\cite{MB-NMNE}.

\begin{figure}
\begin{center}
\includegraphics[width=0.6\textwidth,angle=90]
{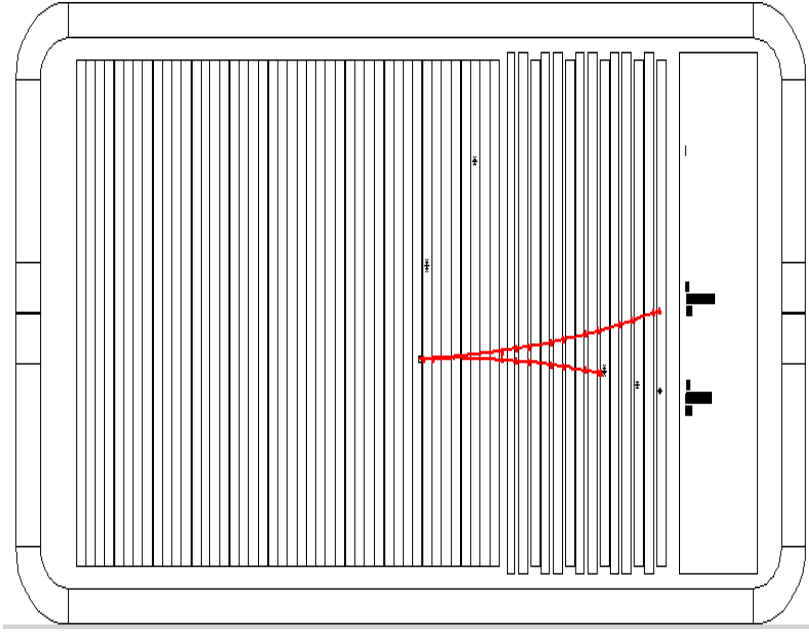}
\caption{Schematic of the DC tracker with
a single  $\gamma$-conversion event in NOMAD in 
Y (vertical) vs Z (horizontal) view.
The energy and angle with respect to the neutrino direction
of the \gam\ are 3.1~GeV and 0.095~Radians. 
}
\label{fig-evt1v01}
\end{center}
\end{figure}

The NOMAD apparatus, described in~\cite{NOMAD-NIM},  
was composed of several sub-detectors. The active 
target comprised 132 planes of $3 \times 3$~m$^2$ drift chambers (DC)    
with an average density similar to that of  liquid 
hydrogen (0.1~gm/cm$^3$). 
On average, the equivalent material in the DC 
encountered by 
particles produced in a $\nu$-interaction  
was about half of a radiation  length 
and a quarter of a hadronic interaction length ($\lambda$).  
The fiducial mass of the NOMAD DC-target,  2.7 tons, was  
composed  primarily of carbon (64\%), oxygen (22\%), nitrogen (6\%), 
and hydrogen (5\%) yielding an effective atomic number, 
\A=12.8, similar to carbon.
Downstream of the DC, there were nine modules of transition radiation 
detectors (TRD), followed by a preshower (PRS) and a lead-glass 
electromagnetic calorimeter (ECAL). 
The ensemble of DC, TRD, and PRS/ECAL was placed within 
a dipole magnet providing a 0.4~T magnetic field orthogonal 
to the neutrino beam line. 
Two planes of scintillation counters, $T_1$ and $T_2$, 
positioned upstream and downstream of the TRD, 
provided the trigger in combination with an 
anti-coincidence signal, ${\overline V}$, 
from the veto counter upstream and outside the magnet. 
Downstream  of the magnet was a hadron calorimeter, 
followed by two muon-stations each comprising large area 
drift chambers and separated by an iron filter. 
The two stations, placed at 8- and 13-$\lambda$'s downstream of 
the ECAL, provided a clean identification of the muons. 
The schematic of the DC-tracker with a $\gamma$-conversion  
in the Y-Z view is shown in Figure~\ref{fig-evt1v01}.
The charged tracks in the DC were measured with an 
approximate  momentum ($p$)  resolution of  
${\rm {\sigma_p/p = 0.05/\sqrt{L} \oplus 0.008p/\sqrt{L^5} }}$  
($p$ in GeV/$c$ and $L$ in meters) 
with unambiguous charge separation in the energy range of interest. 
The experiment recorded over 1.7 million 
neutrino interactions in the active drift-chamber  (DC) target in 
the range ${\cal {O}}(1) \leq E_\nu \leq 300$~GeV.

\clearpage \newpage

\section{The Exclusive Photon Signature}
\label{sec-sig}

In NOMAD the cleanest signature of a \gam\ 
is  a \sv\ arising from its conversion  
in the DC  into $e^-$ and $e^+$ tracks   
as  shown in Figure~\ref{fig-evt1v01}. 
The \gam-momentum is reconstructed by 
measuring the 3-momenta of the associated $e^-$/$e^+$ tracks in the DC. 
The ECAL measures energy associated with the $e^-$/$e^+$ tracks plus any 
unconverted photons. 
Because the trigger counters are situated at the downstream end 
of the detector, only a fraction of events containing single photon  
will trigger the apparatus, namely  events  
where the photon undergoes conversion in DC and 
the resulting $e^+$ or $e^-$ 
traverses the trigger counters. Events with unconverted 
photons  reaching the ECAL  will not trigger the apparatus.   
For example, about 29\% of the \cohp\ events, containing a $\pi^0$ and 
`nothing else', trigger the apparatus; the 
loss arises from the unconverted photons 
and, among the converted photons,  from 
the $e^-/e^+$ tracks that do not reach 
the downstream trigger counters. 

Single photon events will manifest as a \sv, composed of 
$e^-$ and $e^+$,  with 
no additional energy in the ECAL. 
A converted photon is defined as a \sv\ such that 
the invariant mass of 
the $e^-$ and $e^+$ ($M_{ee}$) is  less than 100~MeV which  
selects the  converted photons with  95\% purity and  97\% efficiency. 
Furthermore, if the single photon is 
approximately collinear with the incident neutrino then the 
collinearity of the photon-track offers an additional kinematic handle. 
Because we do not have a definitive 
simulation for the  single \gam\ process, 
we use one of the photons from the \cohp, 
while ignoring the other photon, 
as a {\it guide} for the signal. 
We define two signal boxes. 
${\rm {Box1}}$ comprises single \gam\ events with negligible   
energy in the ECAL associated with neutral particles 
(\gam, neutron, etc.)  quantified by   
the quantity, 
\pan = $\left [ E_\gamma - E_{Neut} \right ] / \left [ E_\gamma + E_{Neut} \right ]$, 
$ \geq 0.9$, where  $E_\gamma$ is the energy of the  photon derived from 
the $e^-$/$e^+$ tracks  
and $E_{Neut}$ is the ECAL 
energy associated with other neutral particles, i.e.  not associated 
with $e^+$/$e^-$ tracks. (For calculation of 
$E_{Neut}$ see, for example, ~\cite{NOMAD-NMNT}.)
The \pan\ $\geq 0.9$ cut is  model independent. 
${\rm {Box2}}$, a subset of ${\rm {Box1}}$, 
comprises single \gam\  events provided that they are 
approximately collinear with the incident neutrino, 
characterized by the variable, 
{\boldmath $\zeta=E_{\gamma} \left [ 1-\cos(\theta_{\gamma}) \right ] \leq 0.05$}, 
where $E_{\gamma}$ and $\theta_{\gamma}$ 
are the photon's energy and  polar angle with respect to 
the neutrino beam. 
(The variable $\zeta$ has the property that its 
distribution depends weakly on the incident neutrino energy.) 
Approximately 90\% of the photons from \cohp\ have  $\zeta \leq 0.05$. 
Although the $\zeta \leq 0.05$  cut will be 
somewhat model-dependent, at the SPS energies 
($E_\nu \simeq 25$~GeV) most exotic models suggest that the single 
photon will be approximately collinear with the incident neutrino.

The backgrounds  to  the single \gam\  are 
all   interactions  
where all daughter particles,  except one photon, 
evade detection. The background includes   the \cohp\ interaction, 
the deep inelastic $\nu$-NC scattering (NC-DIS),  and 
$\nu$-interactions occurring outside the fiducial volume (OBG).
Note, that with a single converted \gam\  in DC 
the coordinates of the  $\nu$-interaction cannot be ascertained. 
The sample with \pan\ $< 0.9$ is composed of events with a \sv\ and 
neutral energy in the ECAL as expected from a \piz\ decay and  constitutes  
an important control sample. 
All three backgrounds are determined {\it in situ} as described below.

The analysis proceeds as follows.  
First, the sample with a single photon-conversion in DC is selected. 
Second, the three backgrounds to the single photon sample are 
calibrated using data outside ${\rm {Box1}}$.
The analysis leads to an experimentally 
constrained prediction of the background events in 
the signal region, ${\rm {Box1}}$ and ${\rm {Box2}}$. 
Finally, boxes  are opened and the predictions are compared with the 
observed data.

\section{Selection of  Single Photon Events} 
\label{sec-sel}
We select $\nu$-induced events where a single photon converts (\sv) 
within the fiducial volume of  the DC target. 
The analysis uses the entire NOMAD data and the 
associated Monte Carlo (MC) samples described 
in ~\cite{NOMAD-XSEC}. 
The NC-DIS sample,  defined by requiring that 
the generated invariant hadronic mass squared (${\rm {W^2}}$) 
be $\geq 1.96$~GeV$^2$,  
is normalized to $0.53 \times 10^6$ 
events which corresponds to 37\% of the \nm-CC. The normalization of the 
NC-Resonance (${\rm {W^2 < 1.96}}$) 
sample is set at 3.5\% of the NC-DIS. 
The \cohp\ interaction is simulated using the 
Rein-Sehgal (RS) model~\cite{RS} which 
agrees well with our measurement~\cite{NOMAD-CohPi0}. 

The MC and data samples are 
subjected to a preselection requiring: 
(a) the presence of a 
converted photon whose reconstructed 
conversion point (${\rm {X}}$, ${\rm {Y}}$, ${\rm {Z}}$)   
be within the fiducial volume, 
${\rm {|X,(Y-5)|\leq 130}}$~cm and \\
${\rm {Z_{Min} \leq Z \leq 405}}$~cm where 
${\rm {Z_{Min}}}$ is 5 and 35~cm for the two configurations of the 
detector composing more than 95\% of the NOMAD data;  
(b) no reconstructed muon  ($\mu$ID-veto) and a single 
photon-induced \sv; 
(c) the ${\rm {X}}$ and ${\rm {Y}}$ coordinate of the  \gam-vector 
when extrapolated back to the upstream most DC at ${\rm {Z_{Min}}}$ 
be within the transverse fiducial volume, ${\rm { |X,Y-5|_{PROJ} \leq130cm}}$,   
largely eliminating   $\gamma$'s that enter from the sides;  
and (d) the energy of the $e^-e^+$ pair be $\geq 1.5$~GeV 
which reduces the NC-DIS and OBG background while having 
a small effect on the \cohp\ sample which also serves  as a guide  
for the single $\gamma$ signal.  
The preselection reduces the NC-DIS and data samples by 
more than a factor of one hundred. 
It is noteworthy that only a small fraction of the NC-Res pass the selection 
since most of the  photons, coming from $\pi^0$ emitted at large 
angles, either fail to trigger the apparatus or have energies below 
$1.5$~GeV. Given the paucity of the NC-Resonance events, 
it  has been added to the NC-DIS sample. 
Finally a negligible fraction of \nm-CC ($<10^{-5}$) 
pass the selection; consequently the CC sample has been ignored in this analysis. 
The preselection is presented in Table~\ref{tab-presel}.

%
%
\begin{table}\centering
\begin{tabular}{||c||c|c|c||c||}
\hline
Cut                                                            &  \cohp\  & NC-Res   &  NC-DIS MC  &  Data    
\\ \hline
Fiducial Cut                                              &   4,900  & 20,000  &  530,000  & 4,018,980 \\

$\mu$ID-veto \& Single $\gamma$-Induced \sv\   
                                                                  &   819    &   306    &    4,330   &     34,062    \\
($M_{ee} \leq 100$~MeV)                         &             &              &               &  \\ 

${\rm {|X,(Y-5)|_{PROJ}\leq130}}$~cm     &     719 &   138     &    3,076    &    10,547 \\ 

$E_{ee} \geq 1.5$ GeV                               &   516  &     69     &  1,846    &      4,543 \\ 
                                                                    &           &              &               &          \\ \hline
\hline
\end{tabular}
\caption{Preselection of events with a single, converted \gam\  
in \cohp, NC-Resonance, 
and NC-DIS MC  samples, and data. 
The MC samples, having a much larger statistics than data, 
are normalized to the expectation as described in the text. 
The ${\rm { |(X,Y-5)|_{PROJ} }}$ refer to X- and Y-position of the \gam-vector 
when projected to the most upstream DC (${\rm { Z_{Min} }}$). } 
\label{tab-presel}
\end{table}

The final event selection follows the preselection 
with more stringent requirements. 
The vertex coordinates of \sv\ are required to be within  \\
${\rm {|X,(Y-5)|\leq 120}}$~cm and ${\rm {Z_{Min} \leq Z \leq 405}}$~cm.  
Two additional cuts  (Clean-\sv) are imposed to 
reduce outside background by requiring 
that there be no tracks upstream of the photon conversion  
and that there be no hits associated with the 
tracks composing the converted-\gam\  in the most upstream DC.
Finally, the $M_{ee}$ cut is tightened to $\leq 50$~MeV which increases 
the photon conversion purity to $\geq 98\%$ while reducing the efficiency to 93\%. 
Table~\ref{tab-sel} summarizes the selection of events in 
the MC  samples.
The preselected data are subjected to identical cuts. 
Distributions of the X, Y, and Z coordinates of 
the \sgam\ vertex in Data and the corresponding MC prediction,  
composed of \cohp\, OBG, NC-DIS, agree. 
Only the control sample with \pan $< 0.9$ is 
examined to check and constrain the background prediction. 
The calibration of the three backgrounds is presented next.

\begin{table}\centering
{\small{

\begin{tabular}{||c||c|c||}
\hline
Cut                                &  \cohp-RS  &  NC-DIS $\oplus$ Resonance   \\ 
\hline \hline      
Start                              &   516         &      1915 \\ 

Tighter Fid-Cuts            &        483   &       1775     \\

$M_{ee}\leq50$~MeV  \& Clean-\sv\    &  386   &   400   \\

\hline \hline      

\end{tabular}
\caption{Final selection of single $\gamma$ Events in the MC Samples:  
The MC samples have been normalized as presented in Section~\ref{sec-sel}.} 
\label{tab-sel}
}}
\end{table}

\section{Constraining the Backgrounds and the  \pan\ $< 0.9$ Sample  } 
\label{sec-back}

The prediction of the backgrounds to the single \gam\ sample  is data driven. 
The estimation of  the \cohp\ induced \sgam\ 
needs to be based on the observed \dv\ sample 
where both photons convert in the DC. 
Furthermore, Monte Carlo simulations can neither reliably provide 
the OBG  nor the  NC-DIS contribution to the \sgam\ samle. 
These backgrounds need to be determined using 
data themselves.

First, we present the calibration of the 
\cohp\ contribution to the \sgam\  sample because 
it is the simplest. The analysis of the \dv\ sample, where 
two photons and nothing else is observed, 
provides a measurement of the  \cohp\ process. It  
yields a rate with respect to the inclusive \nm-CC,  
$\left [ 3.21 \pm 0.46(stat \oplus syst)  \right ] \times 10^{-3}$~\cite{NOMAD-CohPi0}. 
The measured kinematic distributions of the \tgam\ data 
are consistent with the RS-\cohp\ model. 
Consequently we use the simulated RS-\cohp\  events,  normalize   
it to the measured cross-section,  subject these to 
the \sgam\ analysis, and obtain  
the \cohp\ contribution to the \sgam\ sample. 
The analysis yields a normalized prediction of \cohp-induced \sgam\ 
with a 14.5\% precision, shown in Table~\ref{tab-FinalSel}.

Next, we present the calibration of the OBG contribution to the \sgam\ sample. 
This, too, is accomplished using the \dv\ sample as in~\cite{NOMAD-CohPi0}. 
The two reconstructed photon momentum vectors enable one to 
determine the $\nu$-interaction vertex by 
extrapolating the vectors upstream and finding 
the coordinates of their distance of closest approach (DCA).
The procedure defines the DCA-vertex with 
coordinates denoted as DCA-X, DCA-Y, and DCA-Z.  
The DCA-vertex resolution is well understood. 
Among the \tgam\ sample 169 events,  out of 550 events 
(see Table~\ref{tab-DCA-Z}), 
have a DCA vertex of origin upstream of the ${\rm {Z_{Min}}}$  
providing a calibration for the OBG-\sgam\ as follows. 
A different data sample, OBG-Data, which includes 
charged tracks,  is selected 
in which the vertex reconstructed using these charged tracks 
is upstream of the fiducial limit (${\rm {Z \leq Z_{Min}}}$). 
In this sample, the charged tracks are  ignored and the events 
subjected to the \tgam\ analysis (see ~\cite{NOMAD-CohPi0}). 
This OBG-Data sample with \tgam\  yields 
927 events with  {\bf DCA-Z}${\rm {\leq Z_{Min}}}$ and 
451 events with {\bf DCA-Z}${\rm {> Z_{Min}}}$ as shown in Table~\ref{tab-DCA-Z}. 
The number of \tgam\ events with {\bf DCA-Z}${\rm {> Z_{Min}}}$, originating from events 
occurring at ${\rm {Z\leq Z_{Min}}}$ but producing no visible tracks,  
is then calculated as $(451/927)\times 169 = 82.2 \pm 6.1$. 
This constitutes the OBG background to the \tgam\ sample. 
The so normalized  OBG sample is then 
subjected to the \sgam\ analysis 
yielding a calibrated OBG prediction,  
with a 7.7\% error (driven by the statistics of 169 events), 
shown in Table~\ref{tab-FinalSel}.

\begin{table}
 \begin{tabular}{|||c||c|c|||} \hline \hline
                  &   DCA-Z$\ge$Zmin  &  DCA-Z$<$Zmin  \\  \hline  
\tgam\ Data          &        381                   &      169       \\
\tgam\ OBG-Data  &        451                  &      927       \\ \hline \hline 
\end{tabular}
\caption{Events passing and failing the Z-cut in \tgam\ samples in 
Data and OBG-Data}
\label{tab-DCA-Z}
\end{table}

\begin{figure}
\begin{center}
\includegraphics[width=0.8\textwidth]
{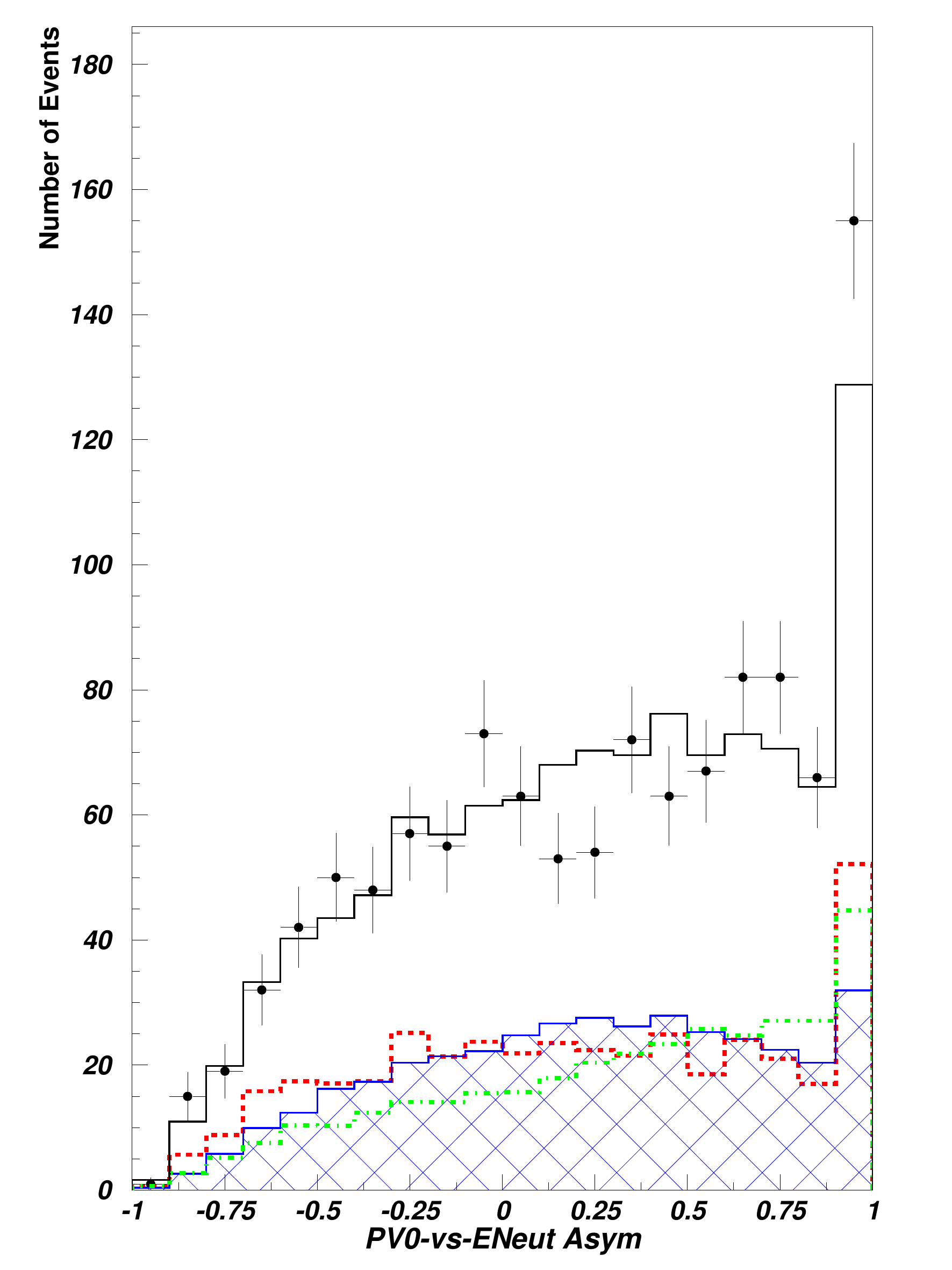}
\caption{Comparison of \pan, the energy asymmetry between 
the \sgam-momentum and the neutral energy in the ECAL, between Data (Symbol) 
and prediction:  
\cohp\ (Blue-hatched), OBG (Green-dash-dot),  
NC-DIS (Red-dash), and total (Black-histogram). 
The signal region, \pan $\geq 0.9$, is not looked at till the analysis is complete.}
\label{fig-Full-PAN}
\end{center}
\end{figure}

\begin{figure}
\begin{center}
\includegraphics[width=0.9\textwidth]
{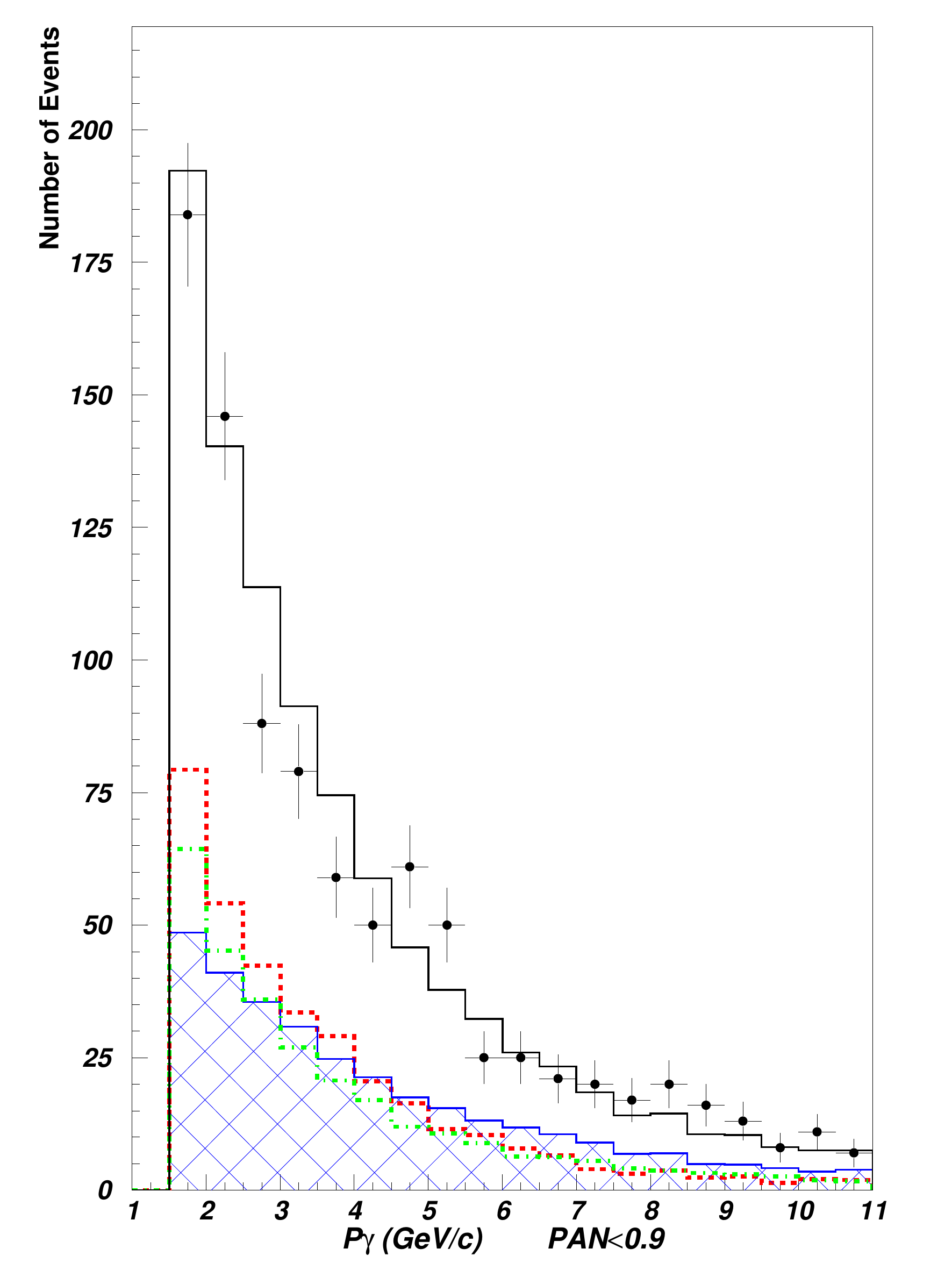} 
\caption{Comparison of $P_\gamma$  between 
data and MC in \pan $< 0.9$ region.}
\label{fig-BkgPan-Pg}
\end{center}
\end{figure}

\begin{figure}
\begin{center}
\includegraphics[width=0.8\textwidth]
{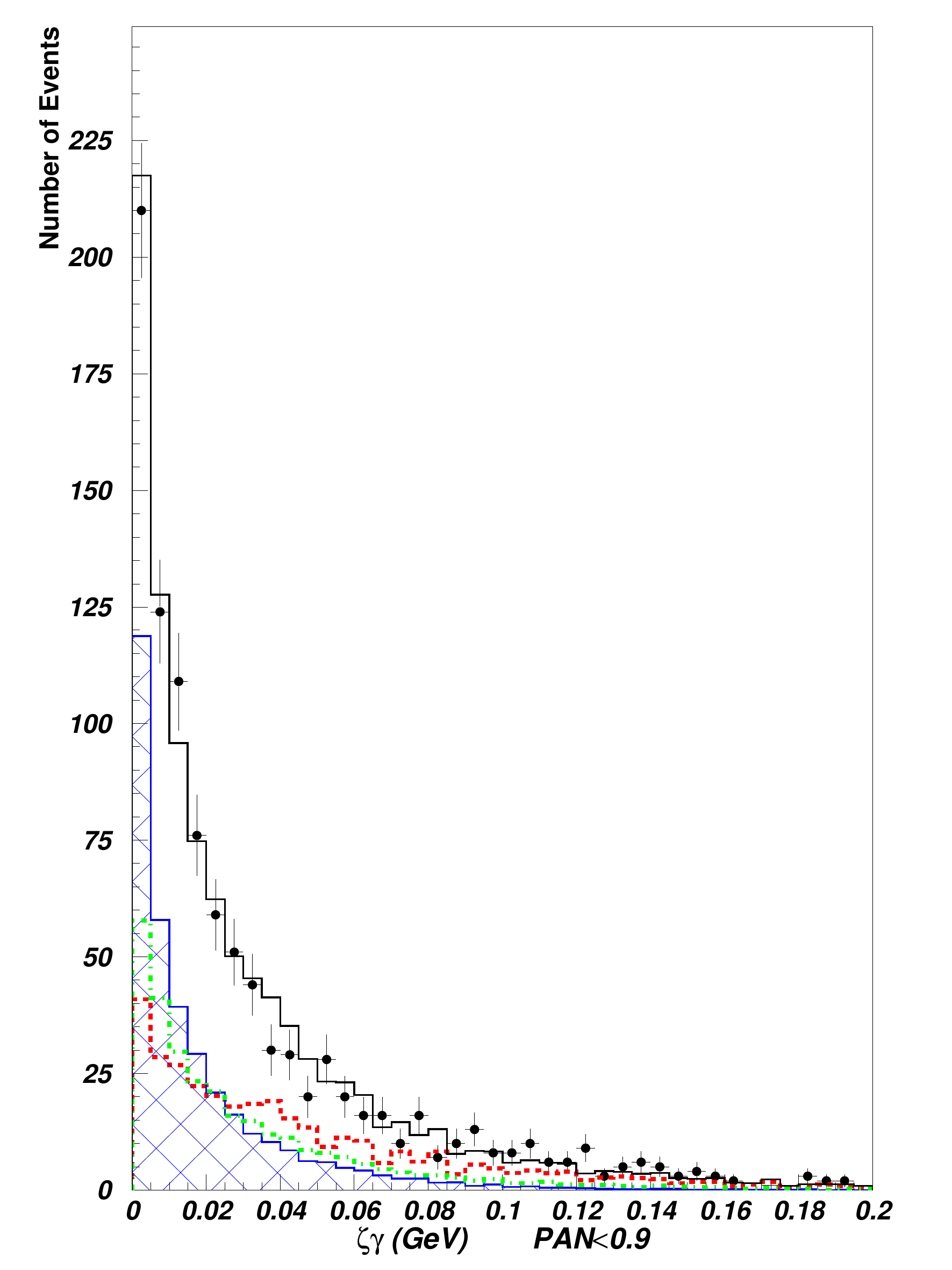}
\caption{Comparison of  $\zeta_\gamma$ distribution of the \sgam\ sample with 
\pan $< 0.9$ in data and MC. The $\zeta_\gamma > 0.05$ region is 
used to normalize the NC-DIS.}
\label{fig-BkgPan-Zeta}
\end{center}
\end{figure}

Finally, we present the calibration of the NC-DIS contribution to the \sgam\ sample 
using the control region \pan $< 0.9$ dominated 
by events containing a photon-conversion accompanied by a neutral energy 
cluster in the ECAL. 
Figure~\ref{fig-Full-PAN} shows  the observed \pan\ distribution for 
the full \sgam\ sample although the signal region \pan $\geq 0.9$ is not 
looked at till all backgrounds are finalized.  
The figure evinces agreement between data and the  prediction for \pan $< 0.9$. 
Furthermore, the observed kinematic variables associated with the neutral cluster in 
ECAL are found to be consistent with the prediction. The observed momentum distribution  
of the photon is consistent with the prediction as evidenced by Figure~\ref{fig-BkgPan-Pg}. 
Lastly, Figure~\ref{fig-BkgPan-Zeta} shows that the measured \ztg\ distribution 
agrees with the MC prediction. Because the \piz-induced photons 
in NC-DIS will have a broader \ztg\  than those in \cohp, we use \ztg $> 0.05$ 
to normalize the NC-DIS. The NC-DIS normalization factor is $1.13 \pm 0.14$. 
(This normalization factor has been applied to the NC-DIS in the figures.) 
The logic behind using this control sample to constrain the background 
in the signal region is that \sgam\ events come from 
$\pi^0$-dominated neutral current events where one of the  photons  
evades detection. One concern, however, is that of a possible systematic 
bias when using the \pan $< 0.9$, a  region where  both 
photons are reconstructed in DC and ECAL, i.e. the two photons are 
emitted in the forward direction, to predict the NC-DIS in the 
\pan $\geq 0.9$ where the second photon  misses the ECAL. 
To check this concern, ordinary NC events, where charged tracks define the event, 
 and, thereby, relieving the  forward bias in the \sgam\ reconstruction due 
 to the trigger requirement, are selected in data and MC.  A single \sgam\ is 
 selected downstream of the event. The charged tracks are ignored, the \sgam\ is 
subjected to the current analysis. The Data/MC ratio  is examined 
in $E_\gamma$ and $\zeta_\gamma$ variables.  The check reveals that 
the shapes of these variables in MC are consistent with those of data: 
the ratio is unity within $\pm 5\%$ in the kinematic range, well 
covered by the 12\% error ascribed to the NC-DIS. 
Table~\ref{tab-FinalSel} presents the NC-DIS contribution to the \sgam\ sample.

To sum up, we have  a calibrated prediction of \cohp, OBG, and 
NC-DIS contributions to the \sgam\ sample. Table~\ref{tab-FinalSel}
lists the corresponding errors, all statistical in nature, as determined by the 
respective data control samples. 
Additionally, we estimate the error  
in the \sgam\ reconstruction to be $\leq 2.5\%$, 
which has a negligible effect on this analysis. 
As a final check, we reproduced the results presented in the 
~\cite{NOMAD-HNU}. 
This exercise illustrates  that the current analysis is 
two orders of magnitude less stringent than that in ~\cite{NOMAD-HNU}  
which was focussed on the search for a heavy neutrino coupling to $\nu_\tau$. 

%
\clearpage \newpage

%
\clearpage \newpage

\clearpage \newpage

\section{Results} 
\label{sec-sig}

\begin{figure}
\begin{center}
\includegraphics[width=0.9\textwidth]
{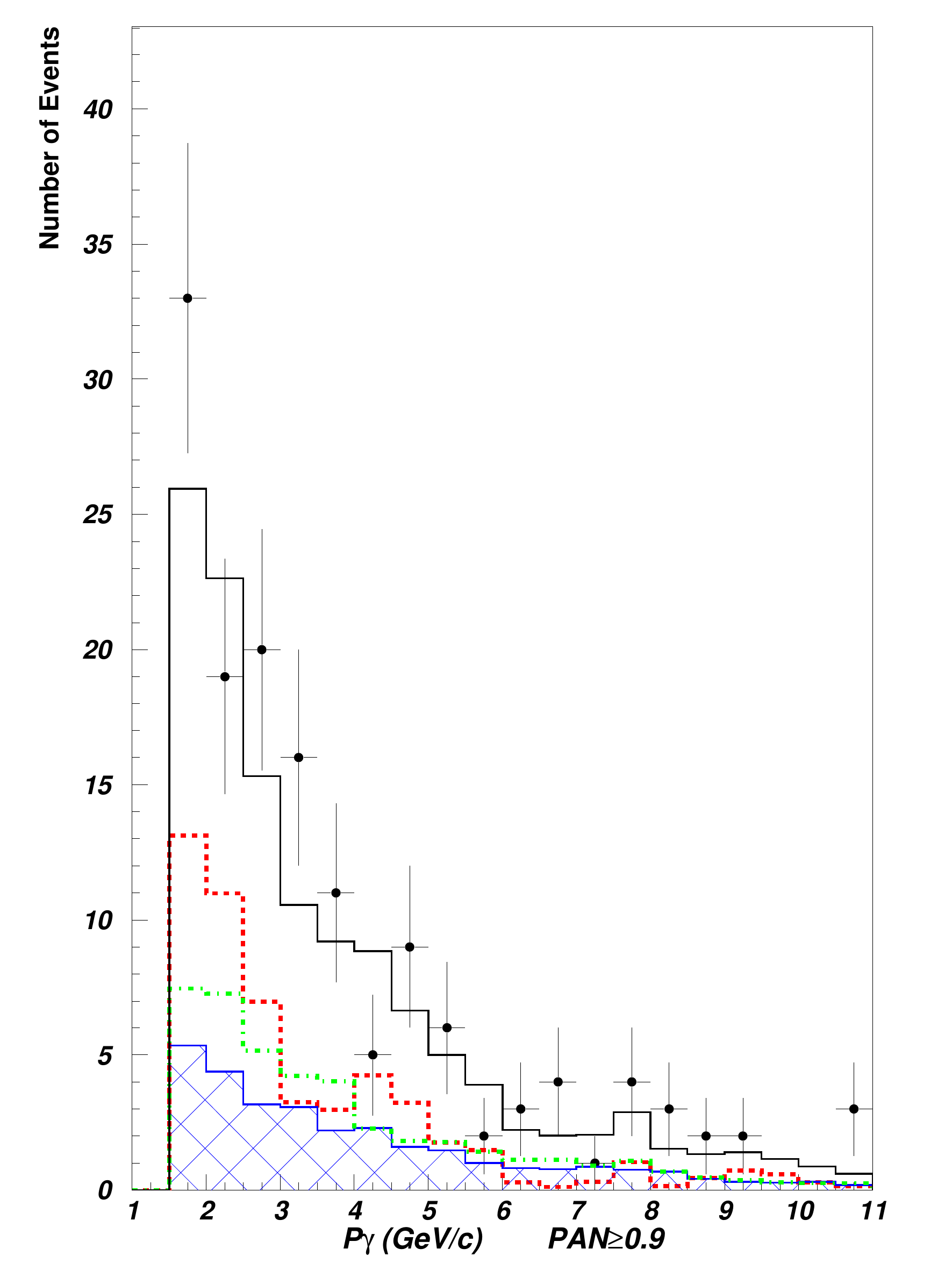} 
\caption{Comparison of $P_\gamma$  between 
data and MC in  ${\rm {Box1}}$,  \pan $\geq 0.9$ region.}
\label{fig-SigPan-Pg}
\end{center}
\end{figure}

\begin{figure}
\begin{center}
\includegraphics[width=0.8\textwidth]
{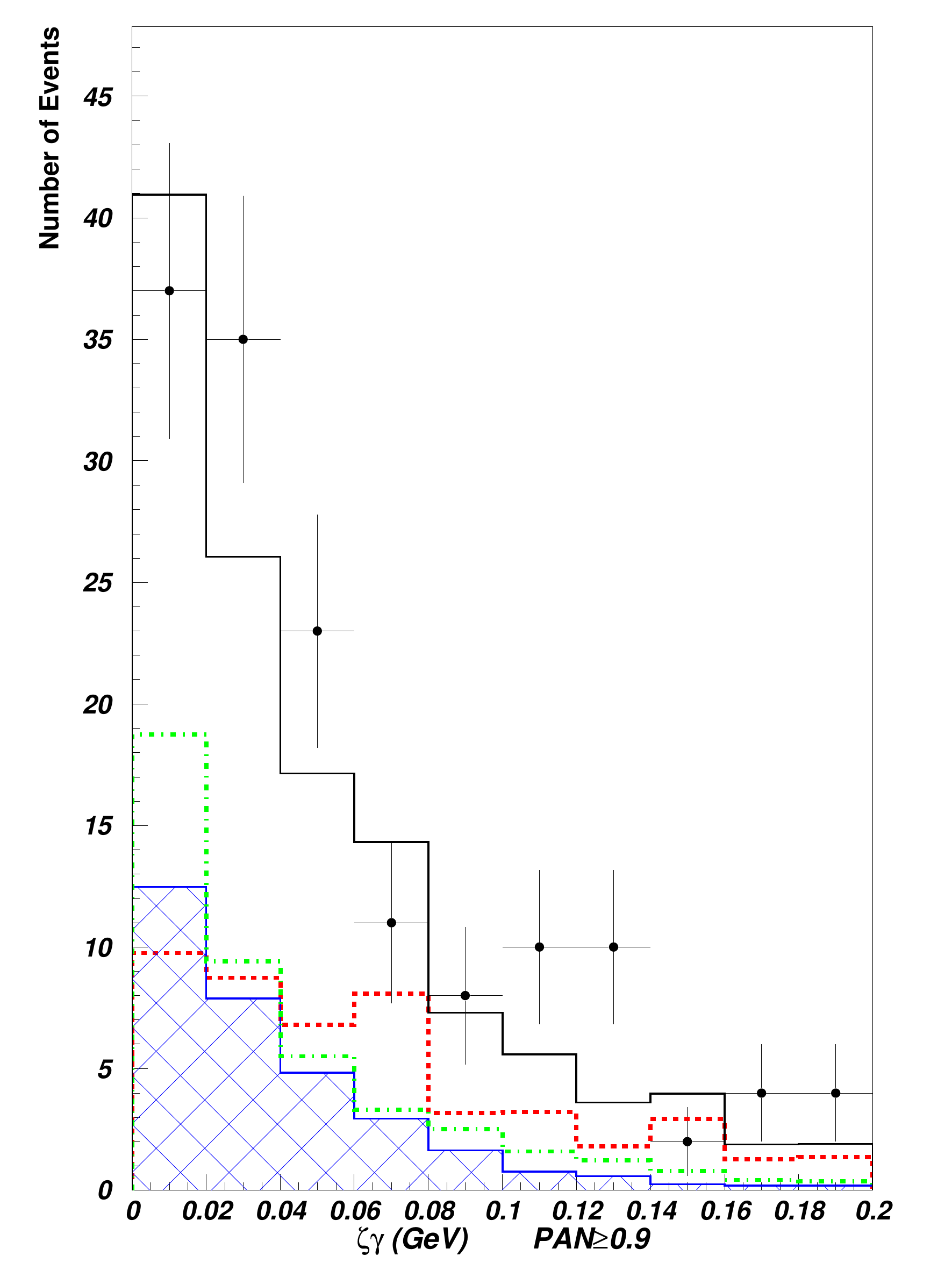} 
\caption{Comparison of  $\zeta_\gamma$ distribution between data and MC 
in ${\rm {Box1}}$,  \pan $\geq 0.9$ region.} 
\label{fig-SigPan-Zeta}
\end{center}
\end{figure}

Figure~\ref{fig-evt1v01} 
shows a representative event in the 
signal region, \pan $\geq 0.9$. 
The first signal-box, ${\rm {Box1}}$,  
with \pan $\geq 0.9$, yields  
155 observed events with a predicted background $129.2 \pm 8.5 \pm 3.3$ events 
yielding an excess of $25.8 \pm 15.5$ events.  
Figure~\ref{fig-SigPan-Pg} and Figure~\ref{fig-SigPan-Zeta} present 
the $P_\gamma$ and $\zeta_\gamma$ comparison between data and MC. 
Absence of significant excess leads us to interpret this as null evidence for 
excess of single photon events. 
Assuming the error on the background-prediction to be Gaussian, 
we derive an upper limit of $<51$ events at 90\% CL in ${\rm {Box1}}$.

\begin{figure}
\begin{center}
\includegraphics[width=0.9\textwidth]
{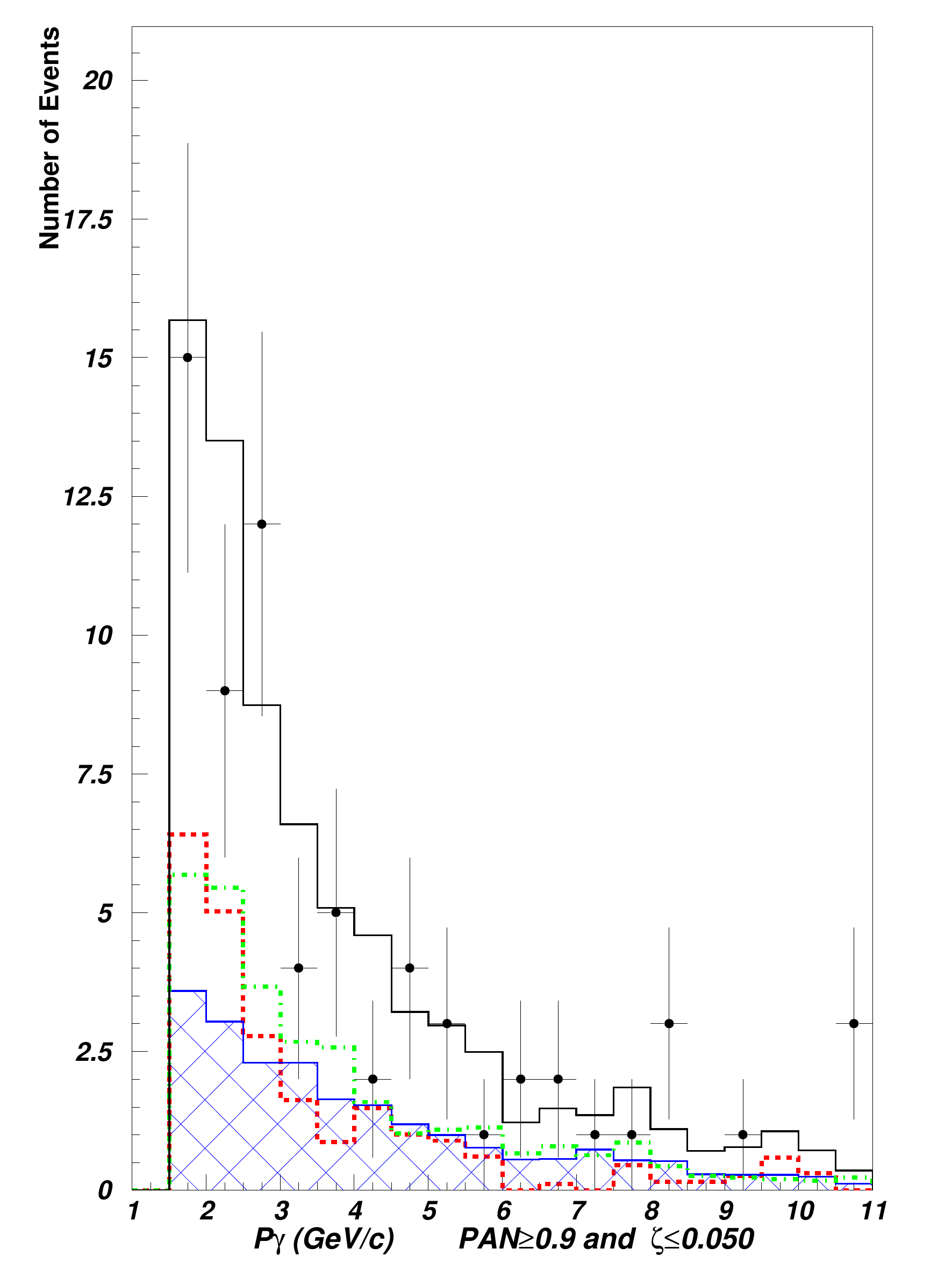} 
\caption{The Momentum of \sgam\ in ${\rm {Box2}}$: 
The observed distribution is consistent with the MC-prediction.}
\label{fig-Sig-Pg}
\end{center}
\end{figure}

\begin{figure}
\begin{center}
\includegraphics[width=0.9\textwidth]
{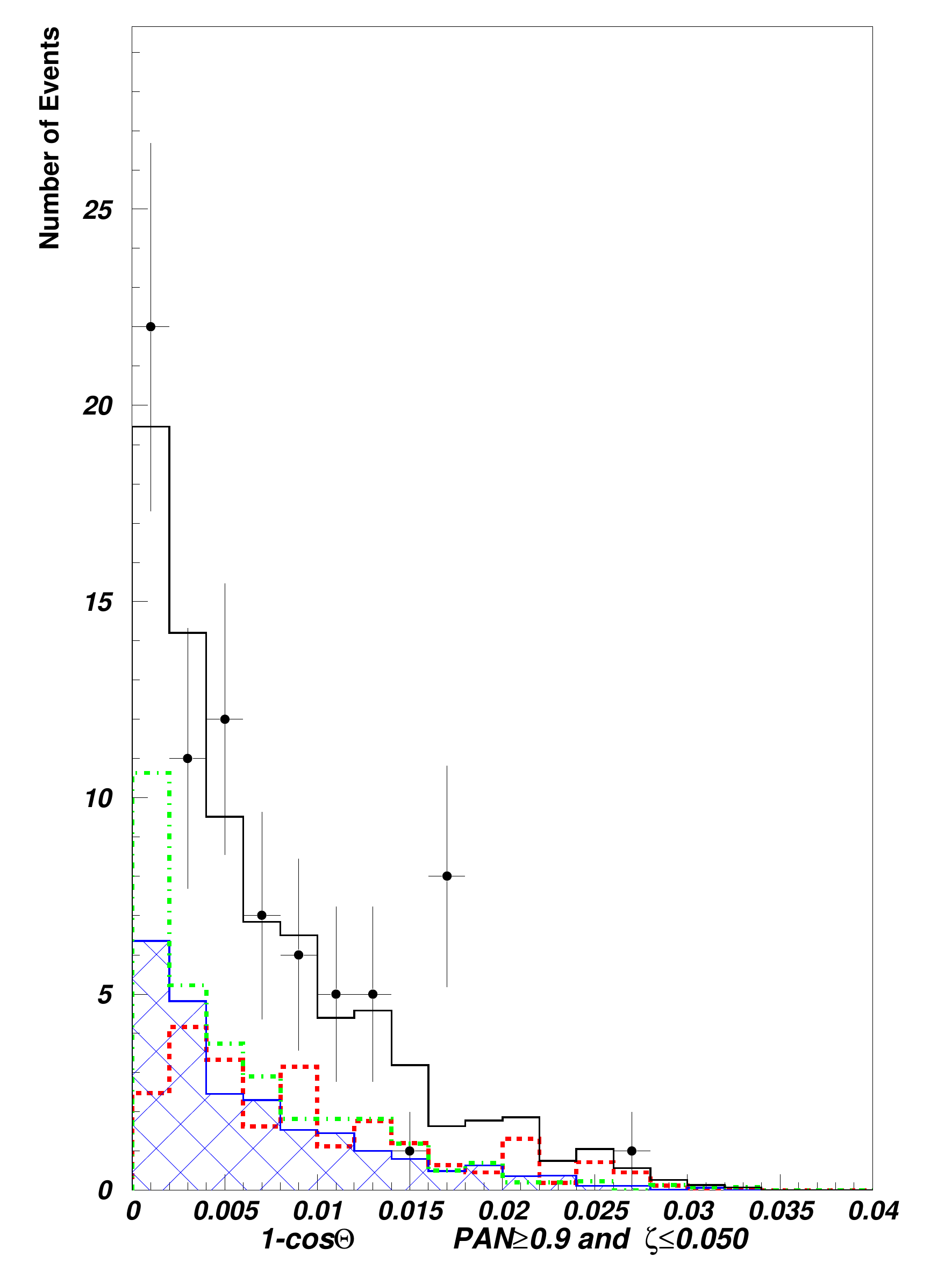}
\caption{Consistent collinearity of \sgam\ in data and MC in the ${\rm {Box2}}$. 
 }
\label{fig-Sig-costh}
\end{center}
\end{figure}

The second signal-box, ${\rm {Box2}}$,  
with \pan $\geq 0.9$ and \ztg $\leq 0.05$, yields 
78 observed events with a predicted background $76.6 \pm 4.9 \pm 1.9$ events 
yielding an excess of $1.4 \pm 10.3$ events.  Events in the  ${\rm {Box2}}$
exhibit kinematic characteristics consistent with the background prediction. 
The vertex-distributions of the \sgam\  events in the signal-region are in 
agreement with the MC prediction; as are, within errors, 
the $P_\gamma$ and $\zeta_\gamma$ distributions,  shown in 
Figure~\ref{fig-Sig-Pg} and Figure~\ref{fig-SigPan-Zeta}. 
In addition, the observed collinearity  (${\cal C}$) 
of the photon matches that of the prediction, see Figure~\ref{fig-Sig-costh}. 
Assuming the error to be Gaussian, we derive an 
upper limit of $<18$ events at 90\% CL.  
Table~\ref{tab-FinalSel} presents the final enumeration. 

\begin{table}\centering
{\small{

\begin{tabular}{|||c||c|c|c|c||c|||}
\hline
Cut                                &  \cohp-RS  &  NC-DIS$\oplus$Res  &  OBG   & Total                   &  Data \\ 
\hline \hline    
\sv\ sample                    &  385.9       &  400.1                         &  341.3  & 1127.3                  & 1149 \\ \hline        
{\bf {MC Error}}             &                   &                                    &             &                             &        \\ 
                                     &    14.5\%    & 12.0\%                        &  7.7\%  &                             &    \\ \hline 
{\bf {Background}}        &                   &                                    &             &                               &        \\ 

\pan\ $< 0.9$              & 353.9                   & 347.7                  &  296.5                   &  998.1 $\pm$69.9$\pm$25.0 
& 994  \\ \hline

{\bf {Signal}}                 &                   &                                    &             &                             &        \\ 

\pan\ $\geq 0.9$          &  32.0          &  52.4                           &  44.8                     & 129.2$\pm$8.5$\pm$3.3   
& 155  \\ 

\pan\ $\geq 0.9$ $\oplus$ \ztg $\leq 0.05$   
                                   & 22.8           &  22.6                            &  31.2                   &    76.6$\pm$4.9$\pm$1.9      
&   78  \\

 \hline \hline 
\end{tabular}
\caption{The \sgam\ Sample: 
Presented are the normalized \cohp, NC-DIS, and OBG predictions 
for the \sgam\ sample. The MC errors of the three components are 
all statistical in nature, as determined by the respective control samples.  
The systematic error due the \sv-reconstruction is shown under the `Total' column. 
Data are shown in the last column.} 
\label{tab-FinalSel}
}}
\end{table}

The only remaining task is to set an upper limit 
on the rate of single $\gamma$ events. To determine the 
signal efficiency, we assume that the `signal' photon has kinematics 
similar to that of one of the photons from the \cohp\ interaction  (the 
other photon is removed from the simulation). Using the RS \cohp-model, 
the signal efficiency for ${\rm {Box1}}$, \pan$\geq 0.9$, is 8.8\%. 
The efficiency for ${\rm {Box2}}$, \pan$\geq 0.9$ and $\zeta \leq 0.05$, 
is 8.0\%. 
The number of fully corrected \nm-CC in the same fiducial volume 
is measured to be $1.44 \times 10^{6}$. We obtain 
the following upper limits on the rate of  single photon events 
in $\nu$-interactions:

\begin{equation} 
\frac { \sigma {\rm { (Single-\gamma) }} }
{\sigma (\nu_\mu {\cal A} \rightarrow \mu^- X)} <  
4.0 \times 10^{-4} {\rm (90\% CL)} 
\label{eq-ccrat1}
\end{equation}

\begin{equation} 
\frac { \sigma {\rm {(Single, Forward-\gamma) }} }
{\sigma (\nu_\mu {\cal A} \rightarrow \mu^- X)} <  
1.6 \times 10^{-4} {\rm (90\% CL)} 
\label{eq-ccrat2}
\end{equation}

\vskip 0.75cm
\noindent
In summary, 
we have presented a search for  single photon 
events in interactions of neutrinos with average energy 
$E_{\nu} \simeq 25$ GeV. 
All relevant backgrounds are constrained using 
data control samples. No significant excess is seen. 
Assuming that the hypothetical signal has kinematics similar 
to those of a photon from the \cohp\ interaction, 
the analysis imposes an upper limit on 
the rate of excess of single photon events 
of $< 4.0 \times 10^{-4}$ per \nm-CC at 90\% CL; 
with an additional soft collinearity cut 
(permitting 90\% of $\gamma$ from \cohp) 
 the limit is $< 1.6 \times 10^{-4}$ per \nm-CC at 90\% CL. 
Following the report on superluminal neutrinos by the OPERA collaboration ~\cite{OPERA}, 
we are conducting a specialized search for  very forward $e^-e^+$ pairs.

%
\clearpage \newpage

\clearpage \newpage

\clearpage \newpage

\clearpage \newpage

\section*{Acknowledgments}
We extend our grateful appreciations to the CERN SPS staff for the magnificent 
performance of the neutrino beam. 
We (CTK and SRM) warmly thank Bill Louis, Richard Hill, Chris Hill, James Jenkins 
and Terry Goldman for many stimulating discussions and insights. 
The experiment was supported 
by the following agencies: 
ARC and DIISR of Australia; IN2P3 and CEA of France, BMBF of 
Germany, INFN of Italy, JINR and INR of Russia, FNSRS of 
Switzerland, DOE, NSF, Sloan, and Cottrell Foundations of 
USA, and VP Research Office of the University of South Carolina.

\end{document}